\documentclass[12pt]{article}
\usepackage[margin=1in]{geometry}% <--- 1 in margin
\usepackage{setspace}

\usepackage{float}
\usepackage{amsfonts}

\usepackage{amssymb}
\usepackage{latexsym}
\usepackage{graphicx}
\usepackage[english]{babel}
\usepackage[font={small}]{caption}

\usepackage{amsfonts}
\usepackage{latexsym}
\usepackage{graphicx}
\usepackage[english]{babel}
\usepackage{tikz}

\begin{document}
\input epsf

\def\p{\partial}
\def\h{{1\over 2}}
\def\be{\begin{equation}}
\def\bea{\begin{eqnarray}}
\def\ee{\end{equation}}
\def\eea{\end{eqnarray}}
\def\d{\partial}
\def\la{\lambda}
\def\eps{\epsilon}
\def\bb{\bigskip}
\def\mm{\medskip}
\newcommand{\dm}{\begin{displaymath}}
\newcommand{\edm}{\end{displaymath}}
\renewcommand{\b}{\tilde{B}}
\newcommand{\gm}{\Gamma}
\newcommand{\ac}[2]{\ensuremath{\{ #1, #2 \}}}
\renewcommand{\ell}{l}
\newcommand{\z}{\ell}
\newcommand{\newsection}[1]{\section{#1} \setcounter{equation}{0}}
\def\bb{$\bullet$}
\def\Qbar{{\bar Q}_1}
\def\QPbar{{\bar Q}_p}

\def\q{\quad}

\def\bn{B_\circ}

\let\a=\alpha \let\b=\beta \let\g=\gamma \let\d=\delta \let\e=\epsilon
\let\c=\chi \let\th=\theta  \let\k=\kappa
\let\l=\lambda \let\m=\mu \let\n=\nu \let\x=\xi \let\r=\rho
\let\s=\sigma \let\t=\tau
\let\vp=\varphi \let\vep=\varepsilon
\let\w=\omega      \let\G=\Gamma \let\D=\Delta \let\Th=\Theta
                     \let\P=\Pi \let\S=\Sigma

\def\h{{1\over 2}}
\def\t{\tilde}
\def\r{\rightarrow}
\def\nn{\nonumber\\}
\let\bm=\bibitem
\def\Kt{{\tilde K}}
\def\b{\bigskip}

\let\p=\partial

\newcommand\blfootnote[1]{%
  \begingroup
  \renewcommand\thefootnote{}\footnote{#1}%
  \addtocounter{footnote}{-1}%
  \endgroup
}

\newcounter{daggerfootnote}
\newcommand*{\daggerfootnote}[1]{%
    \setcounter{daggerfootnote}{\value{footnote}}%
    \renewcommand*{\thefootnote}{\fnsymbol{footnote}}%
    \footnote[2]{#1}%
    \setcounter{footnote}{\value{daggerfootnote}}%
    \renewcommand*{\thefootnote}{\arabic{footnote}}%
    }

\begin{flushright}
%OHSTPY-HEP-T-03-012\\
\end{flushright}
\vspace{20mm}
\begin{center}
{\LARGE The universality of black hole thermodynamics
 }

\vspace{18mm}
{\bf Samir D. Mathur$^{1}$ and Madhur Mehta$^2$
}

\blfootnote{$^{1}$ email: mathur.16@osu.edu }
\blfootnote{$^{2}$ email: mehta.493@osu.edu}

\vspace{4mm}

\b

Department of Physics

 The Ohio State University
 
Columbus,
OH 43210, USA

\b

\vspace{4mm}
% March 31, 2023
\end{center}
\vspace{10mm}
\thispagestyle{empty}
\begin{abstract}

The thermodynamic properties of black holes -- temperature, entropy and radiation rates -- are usually associated with the presence of a horizon. We argue that any Extremely Compact Object (ECO) must have the {\it same} thermodynamic properties. Quantum fields just outside the surface of an ECO have a large negative Casimir energy similar to the Boulware vacuum of black holes. If the thermal radiation emanating from the ECO does not fill the near-surface region at the local Unruh temperature, then we find that no solution of gravity equations is possible. In string theory,  black holes microstates are horizonless quantum objects called  fuzzballs that are expected to have a surface $\sim l_p$ outside $r=2GM$; thus the information puzzle is resolved while preserving the semiclassical thermodynamics of black holes.

\end{abstract}
\vskip 1.0 true in

\newpage
\setcounter{page}{1}

\doublespace

\section{Introduction}

In 1975 Stephen Hawking made a historic discovery: a Schwarzschild black hole of mass $M$ must  radiate at a temperature \cite{hawking}
\be
T_H={1\over 8\pi GM}
\label{one}
\ee
The temperature (\ref{one}) leads to  fascinating thermodynamic  properties of black holes. Black holes must have an entropy \cite{hawking, bek}
\be
S_{bek}={A\over 4G}
\label{two}
\ee
where $A$ is the area of the horizon. Further, the radiation rate is related to the absorption cross-section in the manner required by thermodynamics \cite{hawking}:  quanta in the spherical harmonic $Y_{l,m}$ with energy in the range $(\omega, \omega+d\omega)$ are radiated at a rate
\be
\Gamma_{BH}(l,m,\omega)d\omega= {{\cal P}(l,m,\omega)\over e^{\omega\over T_H}-1} {d\omega\over 2\pi}
\label{three}
\ee
where ${\cal P}(l,m,\omega)$ is the absorption probability for an incoming spherical wave of energy $\omega$ in the spherical harmonic $Y_{l,m}$.

We now see a vexing problem. This thermodynamics appears to rests on Hawking's picture of radiation, where radiation is created from the vacuum around the horizon by the tidal force of gravity. But radiation  emerging from the vacuum carries no information, so it leads to the information paradox \cite{hawking}. Thus it would seem that even if we resolve the puzzle by showing that the hole has a different structure - the structure of a normal body that radiates from its surface --  we would still be left with the problem of explaining why this body should have the thermodynamic properties that we have come to expect from black holes.

In this essay we will show the following:  {\it Any body whose radius is sufficiently close to the radius of the hole, will have the same thermodynamic properties as the semiclassical hole considered by Hawking}.

The above observation will tie up neatly with the fuzzball paradigm which provides a resolution of the information puzzle. In string theory, several classes of  microstates of black holes have been explicitly constructed, and in each case it is found that they have the structure of a normal body with no horizon \cite{fuzzballs}. An entropic argument indicates that the surface of a generic fuzzball should be order  Planck length $l_p$ outside the Schwarzschild radius $2GM$ of the semiclassical hole \cite{ghm}. The argument presented below will then show that any such `Extremely Compact Object' -- ECO for short -- will reproduce the same thermodynamic  properties (\ref{one}),(\ref{two}),(\ref{three}). We define an Extremely Compact Object by the following requirements:

ECO1: The mass as seen from infinity is $M$. 

ECO2: Standard semiclassical physics is a good approximation to the dynamics at $r \ge R_{ECO}$, where $R_{ECO}$ is the radius of our ECO. (In general there will also be a shell-shaped region inside $R_{ECO}$ where semiclassical physics is valid.)

ECO3: The redshift in the semiclassical region reaches a value $O\left ({M\over m_p}\right )$ (this is value of the redshift at distances of order $\sim l_p$ outside the horizon of the Schwarzschild hole). The mass contained in the region $r\le R_{ECO}$ should be $M(R_{ECO})=M-o(M)$. This condition encapsulates the notion that our ECO is `extremely compact'.

ECO4: We assume that causality holds in our theory. Thus for $r\ge R_{ECO}$ (a region where semiclassical physics holds), the mass $M(r)$ contained within $r$ should satisfy
\be
{2GM(r)\over r}<1
\label{four}
\ee
Violation of (\ref{four}) implies that light cones at $r$ point `purely inwards'. Thus violating (\ref{four}) would imply  that the surface of an ECO maintaining a radius $R_{ECO}$ travels faster than the speed of light, something that is not allowed by causality.

Before proceeding, we note a result derived in \cite{israel}. Suppose the black hole is replaced by a thin spherical shell which is supported by its own pressure, and stands a small distance outside its horizon radius. It was argued that such a shell will have to be in equilibrium with the  local Unruh radiation, and that this fact leads to the entropy (\ref{two}) for the shell. In the present article we will be considering a general ECO (i.e., we will not take a particular model like a thin shell), so our arguments will proceed on slightly different lines. But we will also encounter the fact that the ECO will have to be in equilibrium at the local Unruh temperature, and this observation will play a central role in our arguments.

\section{Relating $S$ and $\Gamma$ to $T$}

We start by noting that if the temperature $T[M]$ of our ECO agreed with the black hole temperature (\ref{one}), then the entropy (\ref{two}) and the radiation (\ref{three}) would automatically agree as well. 

First consider the entropy. For any body with a large number of degrees of freedom, we have $TdS=dE$. Thus if we were given that our ECO had an temperature $T[M]$ that matched the black hole temperature (\ref{one}), then we would have 
$dS_{ECO}=T^{-1} dE= (8\pi GM) dM$, which integrates to
\be
S_{ECO}=4\pi GM^2={A\over 4G}
\ee
This reasoning was used to convert Bekenstein's qualitative conjecture $S\sim {A/G}$ \cite{bek} to the precise relation (\ref{two}) after Hawking's discovery of the temperature (\ref{one}); here we are just noting that {\it any} object with the same $T[M]$  as the black hole would yield the same entropy $S[M]$ as the black hole.

Now consider the radiation rate $\Gamma$. For a general object, there is no connection between the radiation rate and the temperature: the radiation rate depends on the detailed size, shape and composition of the object. But the situation is different for an ECO, as we now note.

The metric of the Schwarzschild hole is, in the region $r>2GM$
\be
ds^2=-(1-{2GM\over r}) dt^2 + {dr^2\over 1-{2GM\over r}} + r^2(d\theta^2+\sin^2\theta d\phi^2)
\label{ten}
\ee
In the region close to the horizon, we can choose coordinates
$
s=\sqrt{8GM(r-2GM)}, ~~~ \t t = {t\over 4GM}
%\label{seven}
$ 
 where (\ref{ten}) becomes Rindler space
\be
ds^2\approx -s^2 d\t t^2 + ds^2 + dx_1^2+dx_2^2
\label{six}
\ee
with $x_1, x_2$ describing the tangent space to the angular sphere. 

In the traditional semiclassical picture of the hole, the region around the horizon is locally a patch of Minkowski space, and the quantum state in this patch  is thus close to the Minkowski vacuum $|0\rangle_M$. The coordinates in (\ref{six}) cover the right Rindler wedge of this Minkowski patch. In the Rindler frame, the Minkowski vacuum appears as the Rindler vacuum plus a set of thermal excitations. In the black hole context, this Rindler vacuum is the Boulware vacuum $|0\rangle_B$, and the temperature is $T(s)\approx {2\pi\over s}$, where $s$ is the proper distance outside the horizon. This temperature equals the local Unruh temperature in the near horizon region, and is also the Hawking temperature (\ref{one}) blueshifted to the local orthonormal frame at $s$:  
\be
T_H (-g_{tt})^{-\h} \approx {1\over 8\pi GM} {4GM\over s}={1\over 2\pi s}
\label{tone}
\ee

 \begin{figure}[h]
\begin{center}
\includegraphics[scale=.45]{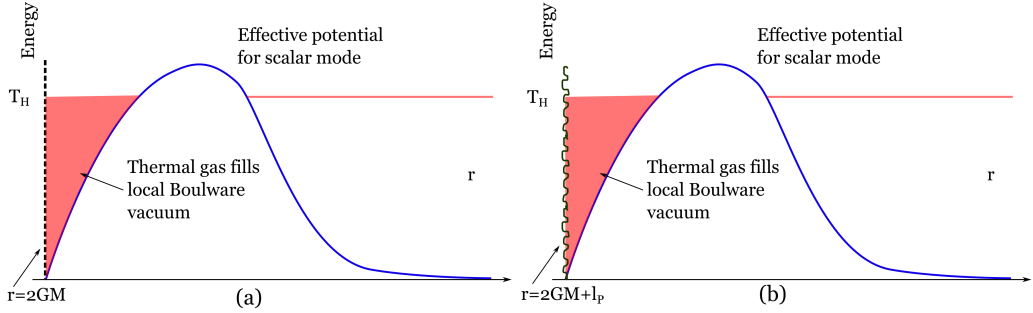}
\end{center}
\caption{(a) The Schwarzschild-frame modes near a black hole fill up to an energy level $E\sim T_H$; leaking through the barrier gives Hawking radiation. (b) An ECO at the same temperature would fill the region just outside $R_{ECO}$ with excitations at the same temperature, giving the same radiation rate.}
\label{fone}
\end{figure}

Let us recall the derivation of Hawking radiation in Schwarzschild coordinates, say for a scalar field $\hat\phi$ satisfying $\square \hat\phi=0$. We expand $\hat \phi$ 
 in modes in the region $r>2GM$
\be
\hat\phi= \sum_{l,m,k} \Big (~   \hat a_{l,m,k}\, f_{l,m,k}(r) Y_{l,m}(\theta,\phi) e^{-i\omega_{l,m,k}\, t}+\hat a^\dagger_{l,m,k}\,  f^*_{l,m,k}(r) Y^*_{l,m}(\theta,\phi) e^{i\omega_{l,m,k}\, t}  ~\Big )
\label{el}
\ee
 The effective potential  felt by these modes $f_{l,m,k}(r)$  is sketched in fig.\ref{fone}(a): there is a barrier separating the region near the horizon from the region at infinity. 
The Hawking radiation rate (\ref{three}) is now obtained as follows: (i) In the near-horizon region the local Minkowski vacuum $|0\rangle_M$ looks like the Boulware vacuum $|0\rangle_B$ plus a thermal gas of excitations $a^\dagger_{l,m,k}$ over this Boulware vacuum; this thermal gas has temperature (\ref{tone}) at a proper distance $s$ from the horizon. 
(ii) These modes tunnel out of the barrier to infinity with a probability ${\cal P}(l,m,k)$, giving rise to Hawking radiation
 (iii) Since the tunneling probability is symmetric for modes going inwards and modes going outwards, this probability ${\cal P}(l,m,k)$ equals the probability that the same mode is absorbed by the hole; this gives the emission rate (\ref{three}). 

But now suppose we replace the horizon by an ECO which has the same temperature as the temperature of a 
black hole. The surface of this ECO will populate the region just outside the surface with quanta of the field $\hat\phi$ with the same temperature (\ref{tone}). The potential barrier is the same as the barrier for the black hole, since the metric in the region outside the ECO is the same as the metric of the hole with the same mass $M$ (fig.\ref{fone}(b)). Thus the radiation $\Gamma_{ECO}(l,m,\omega)$ from the ECO will match the radiation rate (\ref{three}) from the hole; i.e., $\Gamma_{ECO}=\Gamma_{BH}$. 

The crucial point in obtaining this result was that the wavelengths of the quanta in the thermal bath near $r\approx 2GM$ are very short compared to the length scale $\sim GM$ over which the effective potential in fig.\ref{fone} extends. This inequality,  in turn, results from the high redshift condition ECO3, and allows a separation of thermal physics near $r=2GM$ from the dynamics of tunneling through the barrier.

\section{The temperature $T[M]$ of an ECO}

From the above discussion we see that if an ECO had the same $T[M]$ as the black hole, then it would have the same entropy $S[M]$ and the same radiation rate $\Gamma$ as the hole. Thus our issue reduces to showing that an ECO must have the same $T[M]$ as the black hole temperature (\ref{one}).

 \begin{figure}[h]
\begin{center}
\includegraphics[scale=.45]{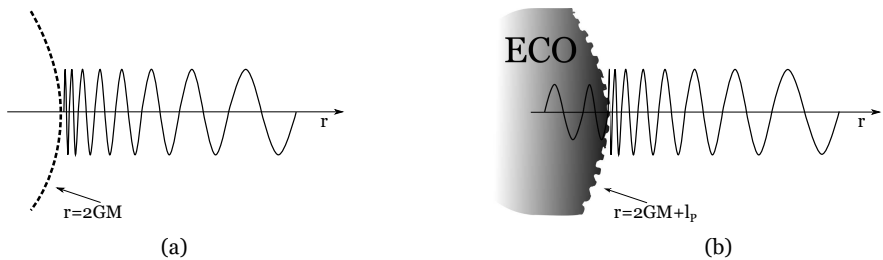}
\end{center}
\caption{(a) The modes in the Schwarzschild metric oscillate infinitely many times before reaching the horizon. (b) The modes in an ECO oscillate in the same way a large number of times before entering the region $r<R_{ECO}$; thus the details of the modes inside the ECO have a negligible impact on the computation of $\langle T_{\mu\nu}\rangle$ in the region outside $R_{ECO}$.}
\label{ftwo}
\end{figure}

Why can  an ECO not have an arbitrary temperature $T_{ECO}[M]$? To answer this, we will have to look at the {\it energy densities} near $r\approx 2GM$. In the black hole, the region around the horizon is a Minkowski vacuum $|0\rangle_M$, so the expectation value of the energy density $\rho=-\langle T_0{}^0\rangle$ is zero, to leading order. In the Schwarzschild frame we find that the energy density of excitations at temperature (\ref{tone}) is
\be
\rho={1\over 480\pi^2} {1\over s^4}
\ee
This energy density is cancelled by the negative Casimir energy density of the  Rindler vacuum $\rho_c=-{1\over 480\pi^2} {1\over s^4}$, giving for the total energy density  $\rho_T=\rho+\rho_c=0$. 

The key point now is that an ECO at zero temperature has the {\it same} negative Casimir energy
 $\rho_{ECO,c}\approx -{1\over 480\pi^2} {1\over s^4}$ at $r\gtrsim R_{ECO}$. To see this, consider the description of the scalar field $\hat\phi$ in the background of the ECO. We expand $\hat \phi$ in modes as
\be
\hat\phi= \sum_{l,m,k} \Big (~   \hat b_{l,m,k}\, g_{l,m,k}(r) Y_{l,m}(\theta,\phi) e^{-i\omega_{l,m,k}\, t}+\hat b^\dagger_{l,m,k}\,  g^*_{l,m,k}(r) Y^*_{l,m}(\theta,\phi) e^{i\omega_{l,m,k}\, t}  ~\Big )
\label{elp}
\ee
and define the vacuum state $|0\rangle_{ECO}$ for $\hat \phi$ by the condition $\hat b_{l,m,k}|0\rangle_{ECO}=0$ for all  $l,m,k$. A wavemode in the black hole geometry is plotted in fig.\ref{ftwo}(a), and a wavemode in the ECO geometry  is plotted in 
fig.\ref{ftwo}(b). The key point is that the wavemode in the ECO has a large number of oscillations in the region $r>R_{ECO}$; the number of these oscillations is $\sim \log({GM\over l_p})$. Thus we can make well-defined wavepackets out of these modes to study the physics at any point $r\gtrsim R_{ECO}$, and so we do not require the form of the modes at $r<R_{ECO}$. 

For the black hole, the computation of the Casimir energy density $\rho_c$ in the Boulware vacuum was carried out in \cite{candelas}. One computes ${}_B\langle 0 | T_0{}^0(s) |0\rangle_B$ using the modes $f_{l,m,k}$ in (\ref{el}). But by the above observation, one finds, to an excellent approximation, the {\it same} result for the ECO using the modes in (\ref{elp}):
\be
\rho_{ECO}=-{}_{ECO}\langle 0 | T_0{}^0(s) |0\rangle_{ECO}\approx  -{}_B\langle 0 | T_0{}^0(s) |0\rangle_B=-{1\over 480\pi^2} {1\over s^4}
\label{tthree}
\ee

 \begin{figure}[h]
\begin{center}
\includegraphics[scale=.25]{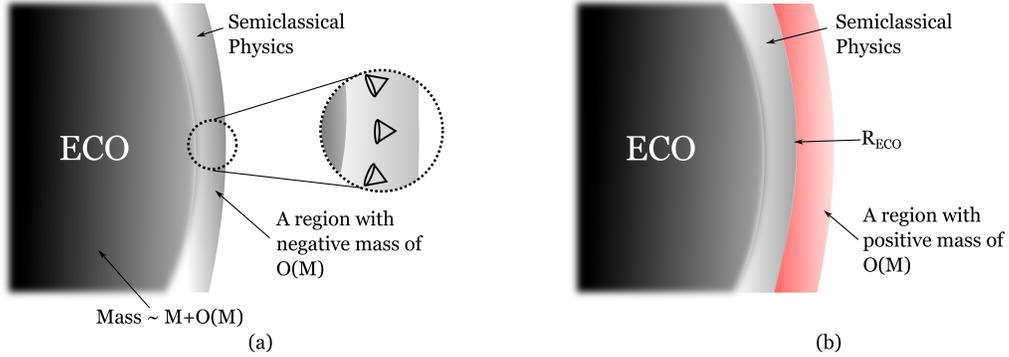}
\end{center}
\caption{(a) For $T_{ECO}<T_H$ the energy in the gray-shaded shell  is $-\beta(T) M$; the light cones in the region just outside the dark shaded region  would have to point inwards, causing the matter there to collapse. (b) For $T_{ECO}>T_H$, the  radiation in the shell just outside $R_{ECO}$ contains an energy $\gamma(T)M$, violating the requirement that there be negligible mass outside $R_{ECO}$.}
\label{fthree}
\end{figure}

Now suppose the ECO was at a temperature $T_{ECO}=0$. Then our state  is just $|0\rangle_{ECO}$. The Casimir energy (\ref{tthree}) implies a negative energy in the semiclassical region just outside $R_{ECO}$. Consider the energy in a  shell shaped region with inner radius a proper distance $\bar s=\mu l_p$ outside $r=2GM$:
\be
E_{shell}\approx 4\pi (2GM)^2\int_{\bar s}^\infty (-{1\over 480\pi^2 s^4}) \left ({s\over 4GM} \right )ds=-{GM\over 240\pi} {1\over    \bar s^2}  =-\big [{1\over 240\pi } {1\over    \mu^2}\big ]  ~M \equiv -\beta \, M
\ee
  We take $\mu = O(1)$; for example we can imagine $R_{ECO}$ is $l_p$ outside the $r=2GM$, and $\bar s= 5 l_p$. Then $\beta = O(1)$ as well. 

Now we see the problem with such an ECO. The energy as seen from infinity is $M$. Thus in the region $r<R_{ECO}$ the energy must be
\be
E_{ECO}=M+\beta M=(1+\beta) M
\ee
But this energy $E_{ECO}$ implies a horizon much outside $R_{ECO}$; in the words, we find a violation of condition ECO4 at $r=R_{ECO}$. Thus such an ECO with temperature $T=0$ cannot exist: ${2M(r)\over r}>1$ at $r=R_{ECO}$ and light cones in the region $r\gtrsim R_{ECO}$ point `inwards', thus forcing the ECO to collapse.

A similar situation holds for any temperature $T_{ECO}<T_H$. The state at temperature $T_{ECO}$ is given by adding a thermal bath of excitations $\hat b^\dagger_{l,m,k}$ to the vacuum $|0\rangle_{ECO}$. The local temperature near the horizon is then 
\be
T_{ECO}(s)={2\pi\over s}{T_{ECO}\over T_H}
\ee
The energy in the shell region is 
\be
E_{shell}\approx -{GM\over 240\pi} {1\over    s_1^2}  \big ( 1- { T_{ECO}^4\over T_H^4} \big )=
-\Big [ {1\over 240\pi} {1\over  \mu^2}  \big ( 1- { T_{ECO}^4\over T_H^4} \big )\Big ] ~M\equiv -\beta(T) M
\ee
with $\beta(T)= O(1)$ again. As in the case $T_{ECO}=0$, such an ECO will also collapse.

Now consider the case $T_{ECO}>T_H$. It turns out that in this case there is no viable solution to the Tolman-Oppenheimer-Volkoff equation describing the thermal radiation in the region  $r>R_{ECO}$. The details of this computation will be presented elsewhere, but the essential idea can be seen by the following approximate argument. By condition ECO3, the redshift reaches $O({M\over m_p})$ at $r=R_{ECO}$. We define a parameter $\bar s$ by writing this redshift as ${4GM\over \bar s}$; then $\bar s\sim l_p$.    By condition ECO2, the region $r\ge R_{ECO}$ is described by standard semiclassical physics. Thus we can find the energy density in this region where the local temperature is $(-g_{tt})^\h T={4GM\over \bar s} T$. The local energy density $\rho$ in this region $r>R_{ECO}$ is then found from this temperature, and the energy in a shell-shaped region at $r>R_{ECO}$ is (writing $\bar s= \mu' l_p$)
\be
E_{shell}\approx 
\Big [ {1\over 240\pi} {1\over  \mu'^2}  \big (  { T_{ECO}^4\over T_H^4}-1  \big )\Big ] ~M\equiv \gamma(T) M
\ee
where $\gamma(T)$ a  positive number of order unity. This is a violation of condition ECO3, which required that the mass outside $r=R_{ECO}$ be $o(M)$; i.e., negligible compared to $M$. (The detailed argument involves including the backreaction of the radiation by solving the Tolman-Oppenheimer-Volkoff equation in the approximation $r\approx 2GM$.)

A more complete analysis (to be presented elsewhere) shows that the above arguments hold for any ECO where the proper distance of the surface at $R_{ECO}$ from the radius $2GM$ is $s\lesssim l_p (M/m_p)^\h$. In $d+1$ spacetime dimensions, this distance is $s\lesssim l_p ( M/m_p)^{2\over (d-2)(d+1)}$.

\section{Summary}

The thermodynamic properties of black holes -- temperature $T_H$, entropy $S_{bek}$ and radiation rates $\Gamma_{BH}$ --  are usually associated to the presence of a horizon; after all Hawking's derivation of $T_H$ had used the separation of geodesics at a horizon and the consequent production of particle pairs. However, this method of radiation leads to information loss \cite{hawking}. This puzzle was made precise in \cite{cern}:  if a semiclassical horizon emerges in any approximation, then one {\it must} have either information loss or remnants. 

Arguments and computations in the fuzzball paradigm of string theory indicate that black hole microstates are horizon-sized quantum objects with a surface  $\sim l_p$ outside  $r=2GM$; these radiate from their surface like normal bodies and there is no information problem. 

The arguments in this article  complement the fuzzball paradigm, by showing that such Extremely Compact Objects would automatically reproduce the thermodynamical properties that we have come to expect from black holes.  
If $T_{ECO}$ differs from $T_H$, then the shell shaped region near $R_{ECO}$ is filled with a Planckian energy density that is either positive or negative. If $T_{ECO}<T_H$, then the energy in this shell is $E_{shell}=-\beta(T)M\sim -M$. The mass  inside $R_{ECO}\approx 2GM$ is then $2GM(1+\beta(T))$, and this mass collapses through its own horizon. If $T_{ECO}>T_H$, then the thermal energy density in a shell just outside $R_{ECO}$ is $E_{shell}\sim \gamma(T) M \sim M$. Then the energy contained within $R_{ECO}$ is not $\approx M$ as required by ECO3; in essence, the thermal radiation around the ECO expands to generate a diffuse object rather than a compact one.

\newpage

 \section*{Acknowledgements}

This work is supported in part by DOE grant DE-SC0011726. We would like to thank Robert Brandenberger for useful discussions. We thank Ted Jacobson for making us aware of \cite{israel}.

\end{document}